\documentclass[aps,pra]{revtex4-2}
\usepackage{graphicx}
\usepackage{subfigure}
\usepackage{xcolor,hyperref}

\begin{document}

\title{Effect of dipole interactions on the properties of an expanding ultracold plasma: A study using quantum mechanical scattering theory.}
\author{Satyam Prakash, Ashok S Vudayagiri}

\affiliation{ School of Physics, University of Hyderabad, Prof.C.R. Rao road, Gachibowli, Hyderabad 500046, \bf{India}}

\begin{abstract}
While generating ultracold plasma (UCP) by photoinization of laser-cooled atoms, only a small fraction of atoms are ionized, and the remaining neutrals interact with the electrons present therein. These interactions, in addition to the Coulomb interactions between ions and electrons, cause phenomena such as ionization of Rydberg atoms and three body recombination, all of which  affect the overall behaviour of the ultracold plasma. We had earlier developed a quantum treatment to analyze these interactions and investigated the ionization of Rydberg atoms in Cesium, which showed good agreement with measured results. We now extend it to other atomic species to investigate Rydberg ionization, and  other effects such as three-body recombination and a resulting additional `quantum pressure' which causes a faster expansion of the UCP. Our results successfully explain experimental observations which were hitherto deemed as `anomalies'. 
\end{abstract}

%\keywords: Rydberg ionisation, ultracold plasma, electron-atom scattering 

%\maketitle

%\begin{document}
	\maketitle
\pagestyle{plain}
\section{Introduction}
Ultracold plasmas (UCP) are partially ionised plasmas, obtained by photoionising a laser-cooled atomic sample, with equilibrium temperatures anywhere between microKelvins to milliKelvins and densities of $\sim 10^9~ cm^{-3}$. These drastically differ from high temperature, high density plasma (HEDP), which have a considerably high temperature (orders of $10^6$ K and upwards) and also high densities ($\sim 10^{21}~cm^{-3}$) \cite{prakash}. An understanding of the underlying physics that control the behaviour of UCP  can be of great interest in probing the high-density plasmas as well as understanding astrophysical systems like white dwarfs and gas giant planets, and the inertial confinement of fusion devices \cite{killian}.

Dynamics of completely ionised HEDPs have been earlier analyzed using the classical kinetic theory, such as Fokker-Planck equations \cite{Spitzer} or relaxation time approach \cite {Blatt, Ziman}. The hydrodynamical equations underlying this approach lead to well-defined analytical formulas for electrical conductivity \cite{Rosmej}. Using a similar kinetic approach, attempts have been made to analyse the dynamics of UCP as well. In  typical analysis, the temperature of ions and atoms is ignored, and the electrons are treated as moving in a mean field created by ions \cite{Pohl,robicheaux}. The expansion of plasma is considered to be adiabatic and is found to thermalise very fast. This would mean that the UCP system is considered to be in equilibrium, giving rise to the spatial correlations between the particles. The simulations of these equations have been applied to explain ultrafast cooling of electrons, expansion of plasma, and the formation of a strongly correlated system on a larger time scale, but do not give adequate information on shorter time scales. Moreover, multiple anomalous behaviors in the UCP system could not be explained while using the kinetic theory approach. For instance, at low electron energies, the plasma expands much faster than expected, which was attributed to the formation of Rydberg atoms in a substantial amount when the plasma expanded \cite{kulin,kilian2}. It was experimentally verified by Kulin et. al. \cite{kulin} by exciting plasma oscillations in a UCP using an RF electric field. They experimentally showed that for $E_e<70 K$ the expansion of the plasma cloud is much faster than expected, largely deviating from the hydrodynamic model. A more improved hydrodynamical model tried to accommodate atomic processes using radiative recombination (RR), dielectronic recombination (DR), and three-body recombination (TBR), which showed a better agreement with the experiments \cite{stevefelt,seaton}. For instance, TBR involves two electrons interacting  with an ion such that one electron gets bound to the atom, and the other electron carries away excess energy. However, combining the TBR  with the radiative redistribution model in plasma at equilibrium predicts a density-independent maximum \cite{stevefelt}, an effect contradicted by the experimental observations of an expanding UCP. The trends observed in multiple experiments \cite{kilian2,vanhaecke} tend towards more deeply bound levels when  the number of ions($N_i$) is increased or the electron energy $E_e$ is decreased. These experimental results were theoretically explained using quantum TBR by Hu et al.\cite{hu}. Moreover, the rate of TBR is predicted to vary with temperature as $T^{-9/2}$ and diverge at low temperatures. Moreover, the rate of TBR predicted to vary with temperature as $T^{(-9/2)}$ as per the classical theory, would diverge at lower temperatures. The same divergence for the TBR as $T^{-9/2}$ was obtained by Hu et al. using the quantum approach. 

Before using the same theoretical model for a UCP system as well as for the HEDP system, it is important to point out the distinct differences between the HEDP system and the UCP system to demarcate the correctness of the theoretical method used for the two systems. The HEDP system is fully ionised, so that the electrons and ions are clearly in electron-ion collision and are well described in a wide parameter range using the kinetic theory approach as discussed by Spitzer and several other authors\cite{Spitzer, Blatt, Ziman}. In a modified approach like that of the linear response theory (LRT), the inclusion of different collision mechanisms is vivid. For instance, Reinholz et al. \cite{reinholz} and Karakhtanov \cite{karakhtanov} have discussed the influence of electron-electron (e-e) collisions within the framework of the Born approximation for an FIP. In contrast to Spitzer’s value, their correction factor is a function of both density and temperature. Now, for a UCP that shows the formation of Rydberg atoms, taking into account the electron-atom collision is not simple within the kinetic theory framework. The formation of a large number of Rydberg atoms during the plasma expansion, as discussed above, gives rise to significant electron-atom interactions. These electron-atom interactions have been considered by Adams et. al. for calculating the momentum transfer cross-section. He used the polarisation potential for electron-atom interaction and combined it with the plasma composition using COMPTRA-04 to obtain the dc conductivity. The results obtained had a better quantitative agreement with experimental results\cite{adams}.
  
In this work, we propose to explain the anomalies shown by UCP using a quantum mechanical approach. Our method takes into account the electron-atom interaction, which is particularly important for partially ionised plasmas like a UCP. In an earlier work, we calculated the scattering cross-sections for Rydberg atom ionization \cite{prakash}, using the potential scattering calculations following the method of Rosmej et al \cite{Rosmej}. The results showed good agreement with experimental results for Cesium atoms \cite{vanhaecke}.  In the present work, we extend the calculations to other alkali atoms and also compute probabilities of three-body recombination. An unexpected result is a 'quantum mechanical pressure' which acts over and above the Coulomb repulsion of ions and electrons, resulting in a faster expansion of the UCP, which has been observed in many experiments.

The screened polarization potential approach also allows us to compute  relevant parameters such as  coupling constant, Debye radius, electron pressure, and plasma frequency during the ultrafast expansion of UCP and show the evolution of these parameters. Our calculations explain how the Rydberg ionisation in a UCP depends on the temperature and density, and also  the anomalous cooling of UCP at lower time scales as reported by Kilian et al \cite{kilian2}. We propose an adequate consideration of the electron-Rydberg atom scattering to explain these experimental observations. 

\section{Formation of Rydberg atoms, TBR, Ionisation in alkali UCP}
An ultracold plasma is formed by the photoionisation of laser-cooled atoms. The laser producing the plasma has a narrow spectral width of ($\sim 10ns$) so that the electrons start with a non-thermal energy distribution and thermalise quickly($<10\mu s$). The time taken to thermalise can be estimated as $\tau_{ee}=1.2 \times 10^{-6}s^4m^{-6}v_e^3/n_e \ln \Lambda$ where $\Lambda = 4\pi \epsilon_o(3k_BT_e\lambda_D/e^2)$ \cite{robicheaux}, $\lambda_D$ is the Debye length, and e is the electronic charge. For electrons at temperature 100 K,  a density $n_e$= $10^{15} m^{-3}$ and  $\ln \Lambda$ = 6.0, the thermalization time is $t_{ee} = 64$ ns. Just before thermalisation, high-temperature electrons boil away, which decreases the temperature and the electron density. Subsequently, the plasma expands and thermalises slowly, which takes up to 50$\mu s$ \cite{kulin}. During the thermalisation, at least 20\% \cite{kilian2} of the free charges recombine through the dominant process of three-body recombination(TBR). In a typical TBR, an electron and an ion recombine in the presence of an extra electron, which serves to conserve the energy and momentum, giving rise to high-lying Rydberg states. The rate of formation of the Rydberg atoms through TBR had been studied as early as the 1960s through classical kinetic approach and two important predictions were made: (1) the rate of TBR scales as $T^{-9/2}$ and (2) The recombination would populate the electrons in the shell level n in such a way that the electron's binding energy is given by $E_{n}=k_BT$. Here, k$_B$ denotes the Boltzmann constant, and T is the electron temperature. 

S. X. Hu investigated the TBR in quantum mechanical formalism by numerically solving the 6-dimensional time-dependent Schrodinger equation: \cite{hu}

\begin{eqnarray}
%\begin{aligned}
i \frac{\partial}{\partial t} \Phi\left(\mathbf{r}_1, \mathbf{r}_2, t\right)&=&  \left[-\frac{1}{2}\left(\Delta_{\mathbf{r}_1}+\Delta_{\mathbf{r}_2}\right)-\frac{1}{r_1}-\frac{1}{r_2}\right. \\ &+&\left.\frac{1}{\left|\mathbf{r}_1 \mathbf{r}_2\right|}~\right] \Phi\left(\mathbf{r}_1, \mathbf{r}_2, t\right) 
%\end{aligned}
\end{eqnarray}

Here, the close-coupling method was used in which the six dimensional wave function given above $\Phi\left(\mathbf{r}_1, \mathbf{r}_2 \mid t\right)$ was expanded in terms of bipolar spherical harmonics $Y_{l_1 l_2}^{L S}\left(\Omega_1, \Omega_2\right), \Phi\left(\mathbf{r}_1, \mathbf{r}_2 \mid t\right)=$ $\sum_{L S} \sum_{l_1 l_2}\left[\Psi_{l_1 l_2}^{(L S)}\left(r_1, r_2 \mid t\right) / r_1 r_2\right] Y_{l_1 l_2}^{L S}\left(\Omega_1, \Omega_2\right)$, for a specific $(L S)$ symmetry. By expanding the Coulomb repulsion term $1 /\left|\mathbf{r}_1-\mathbf{r}_2\right|$ in the form of spherical harmonics, substituting
into the Schrodinger equation and finally integrating over spatial angles $\Omega_1$ and $\Omega_2$, equation (1) can be reduced to the radial variables $r_1$, $r_2$ as given in equation(2). In equation(2), the running index j goes upto the total number of partial waves N. Here, $\hat{T_1}, \hat{T_2}$ , and $\hat{V_c}$  are the diagonal operators which denote the kinetic energies of the electrons and the electron-nucleus Coulomb interaction, respectively. The off-diagonal term $\hat{V}_{j, k}^I\left(r_1, r_2 \mid t\right)$ denotes the Coulomb repulsion between the two electrons. 

\begin{eqnarray}
%\begin{align*}
i \frac{\partial}{\partial t} \Psi_j\left(r_1, r_2 \mid t\right)= & {\left[\hat{T}_1+\hat{T}_2+\hat{V}_c\right] \Psi_j\left(r_1, r_2 \mid t\right) } \\
& +\sum_k \hat{V}_{j, k}^I\left(r_1, r_2 \mid t\right) \Psi_k\left(r_1, r_2 \mid t\right)
%\end{align*}
\end{eqnarray}

To solve equation(2), each electron is described in terms of Gaussian wave function, $G_{k l}(r)=\left[(i)^l /\left(\pi w^2\right)^{1 / 4}\right] \times$ $e^{-\left(r-r_0\right)^2 / 2 w^2} e^{-i k r}$. Here $w$ represents the width associated with the electron wave packet having momentum $k$, and $l$ denotes its azimuthal quantum number. Calculation of the total electron wavefunction is written in the product form given by: $G_{k_1 l_1}$ and $G_{k_2 l_2}: \Psi_j\left(r_1 r_2, t=0\right)=(1 / \sqrt{2})\left[G_{k_1 l_1}\left(r_1\right) G_{k_2 l_2}\left(r_2\right)+\right.$ $\left.(-1)^S G_{k_1 l_1}\left(r_2\right) G_{k_2 l_2}\left(r_1\right)\right]$. S. X. Hu combined the FEDVR (finite-element discrete variable representation) with the RSP(real-space product) algorithm to obtain the TBR probability density as given by equation(3). Here,$\phi_{n l}$ and $\phi_{k_2 l_2}$ are the bound and continuum states respectively and $E_2$ is the energy of the outgoing electron. 

\begin{equation}
P_{n l}\left(E_2\right)=2 \sum_{L S} \sum_{l_2}\left|\int d r_1 \int d r_2 \phi_{n l}^*\left(r_1\right) \phi_{k_2 l_2}^*\left(r_2\right) \Psi_{l l_2}^{(L S)}\left(r_1, r_2, t=t_f\right)\right|^2
\end{equation}

Integrating $P_{n l}\left(E_2\right)$ over $E_2$, we get the probability of the TBR for the specific quantum numbers $n, l$, $P_{n l}=\int P_{n l}\left(E_2\right) d E_2$ and the total probability of recombination for $n$ level is obtained by summing the recombination probability over the angular momentum $l$, given by, $P_n=$ $\sum_l P_{n l}$. Summing over the s-wave contributions, it has been shown[11] that for electron energies lying between $0.05eV<E_e<5eV$, the rate of recombination is maximum for the Rydberg levels for which $E_n \sim k_BT$. It is important to note that for energy lower than 0.05 eV, the quantum mechanical calculations showed deeper bound states than those given by $E_n \sim k_BT$.

\begin{figure}[!h]
	\subfigure(a){\includegraphics[width=7.2 cm]{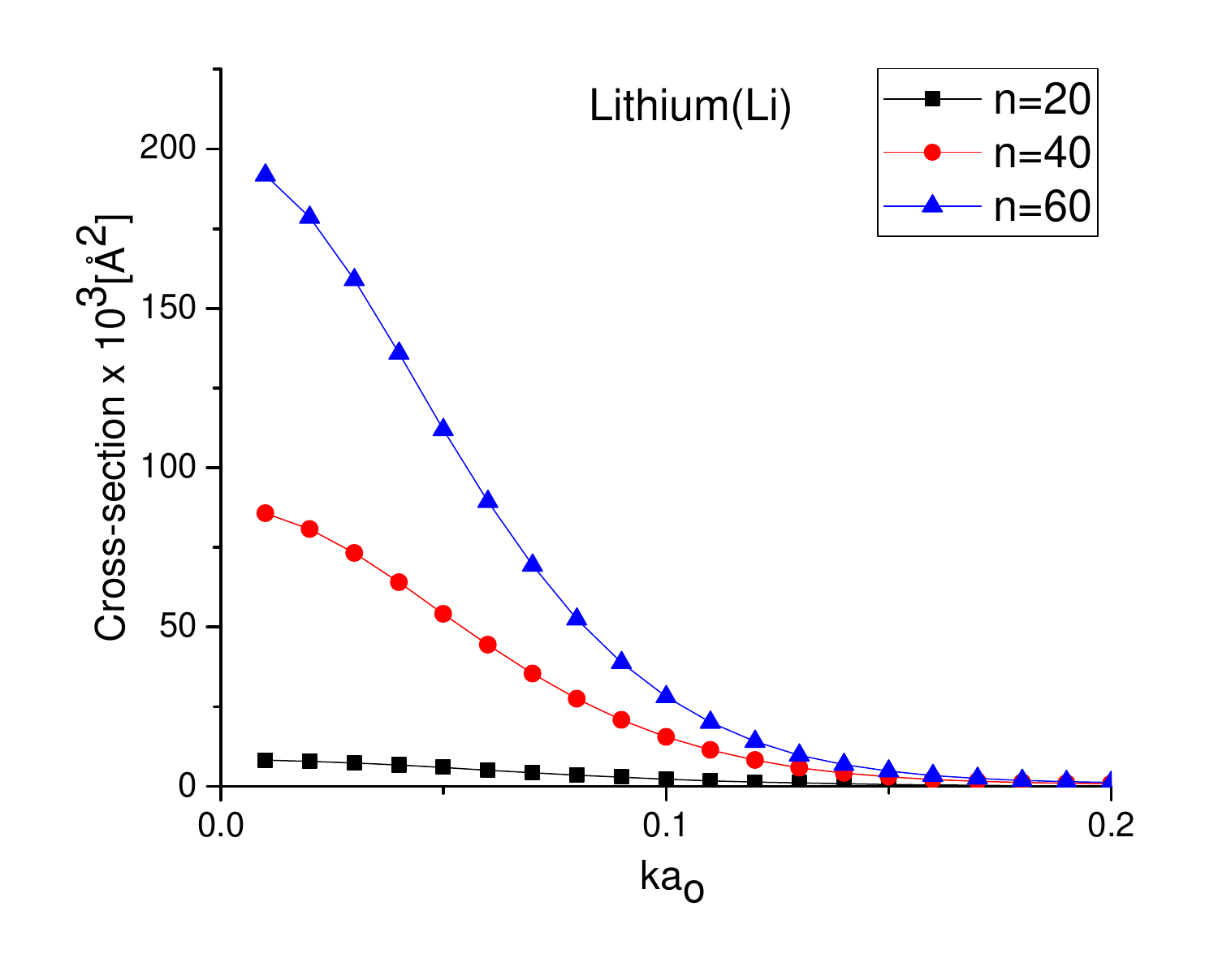}}	
	\subfigure(b){\includegraphics[width=7.2 cm]{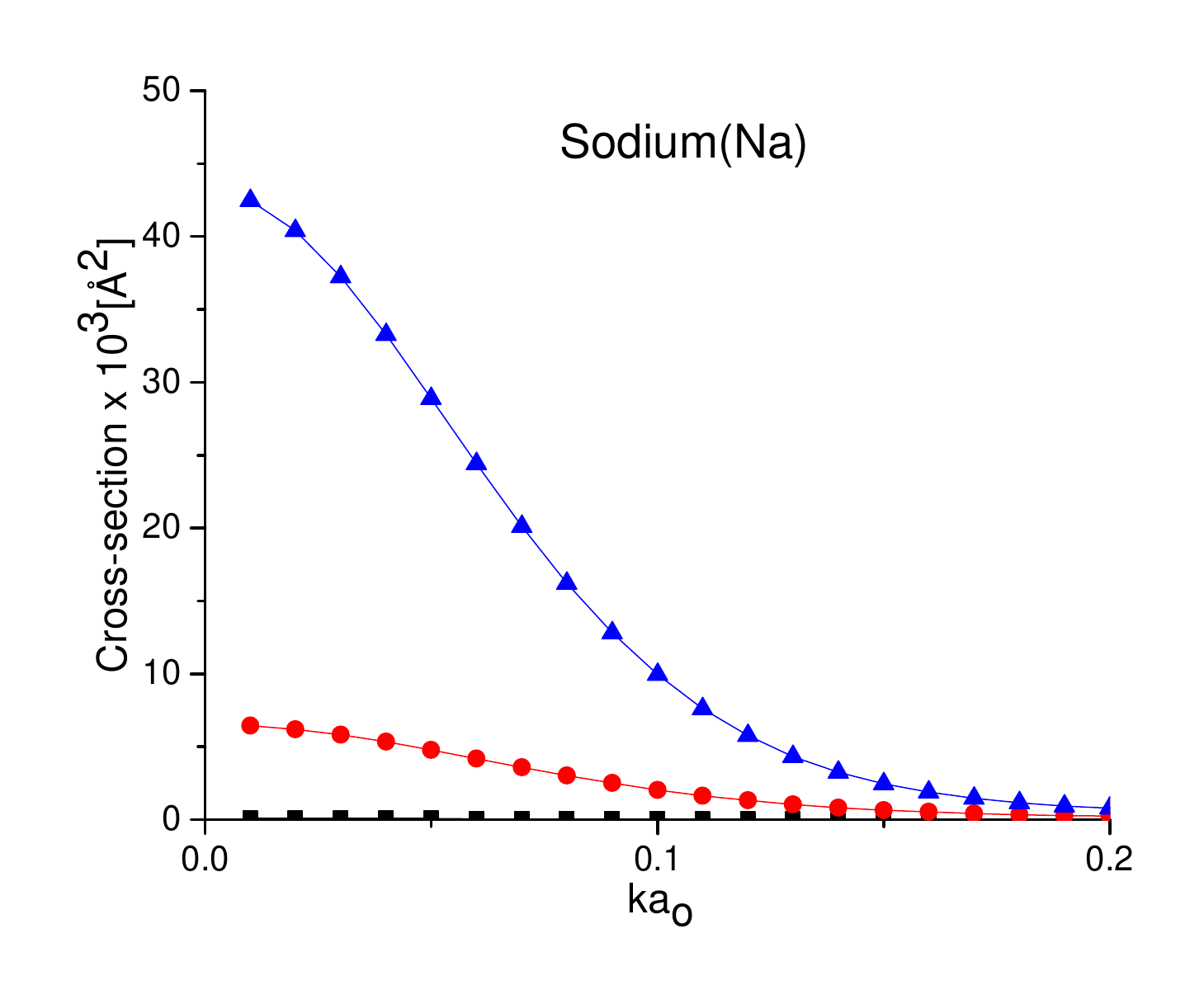}}
	\subfigure(c){\includegraphics[width=7.2 cm]{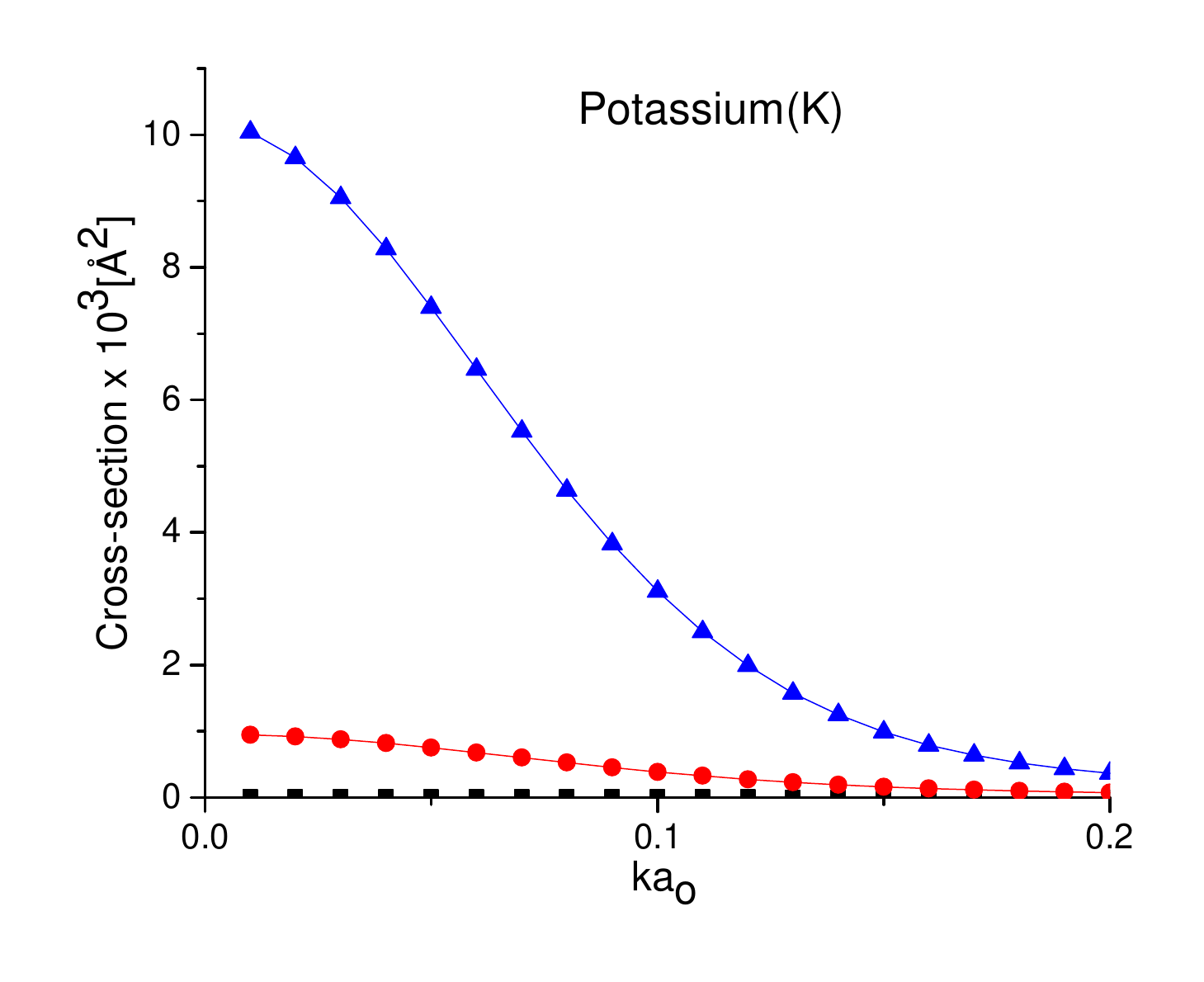}}
	\subfigure(d){\includegraphics[width=7.2 cm]{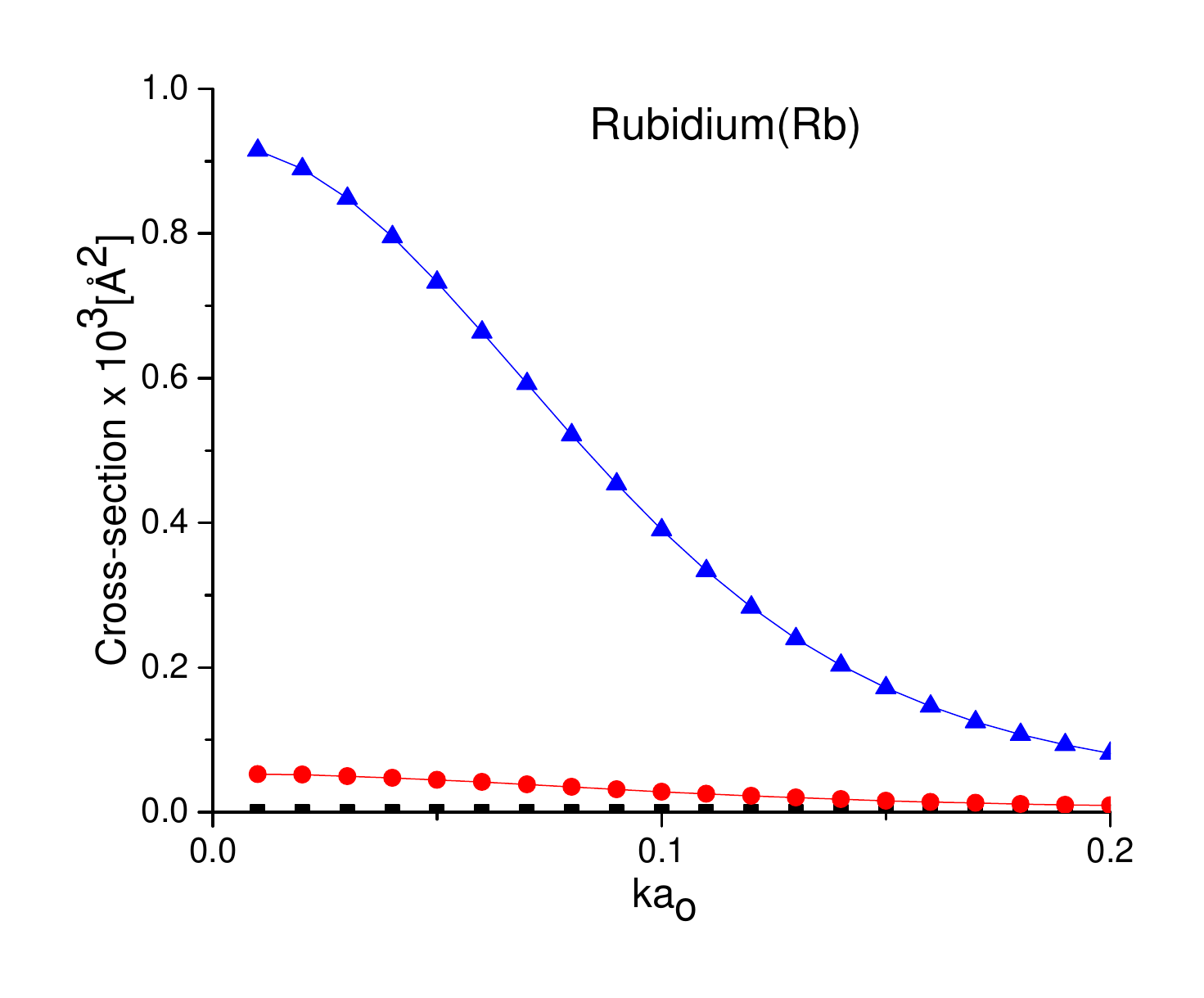}}
	\subfigure(e){\includegraphics[width=7.2 cm]{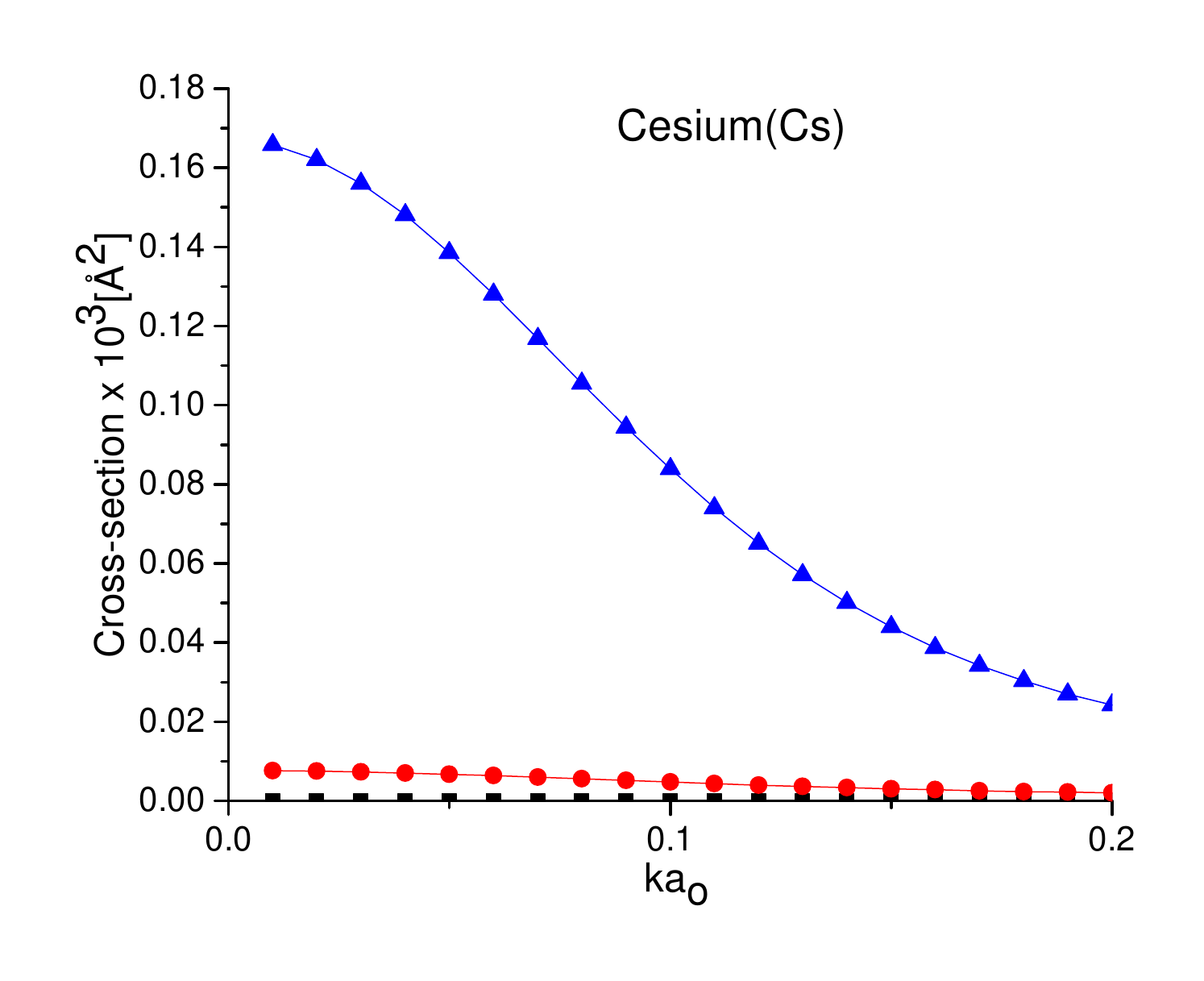}}
	\subfigure(f){\includegraphics[width=7.2 cm]{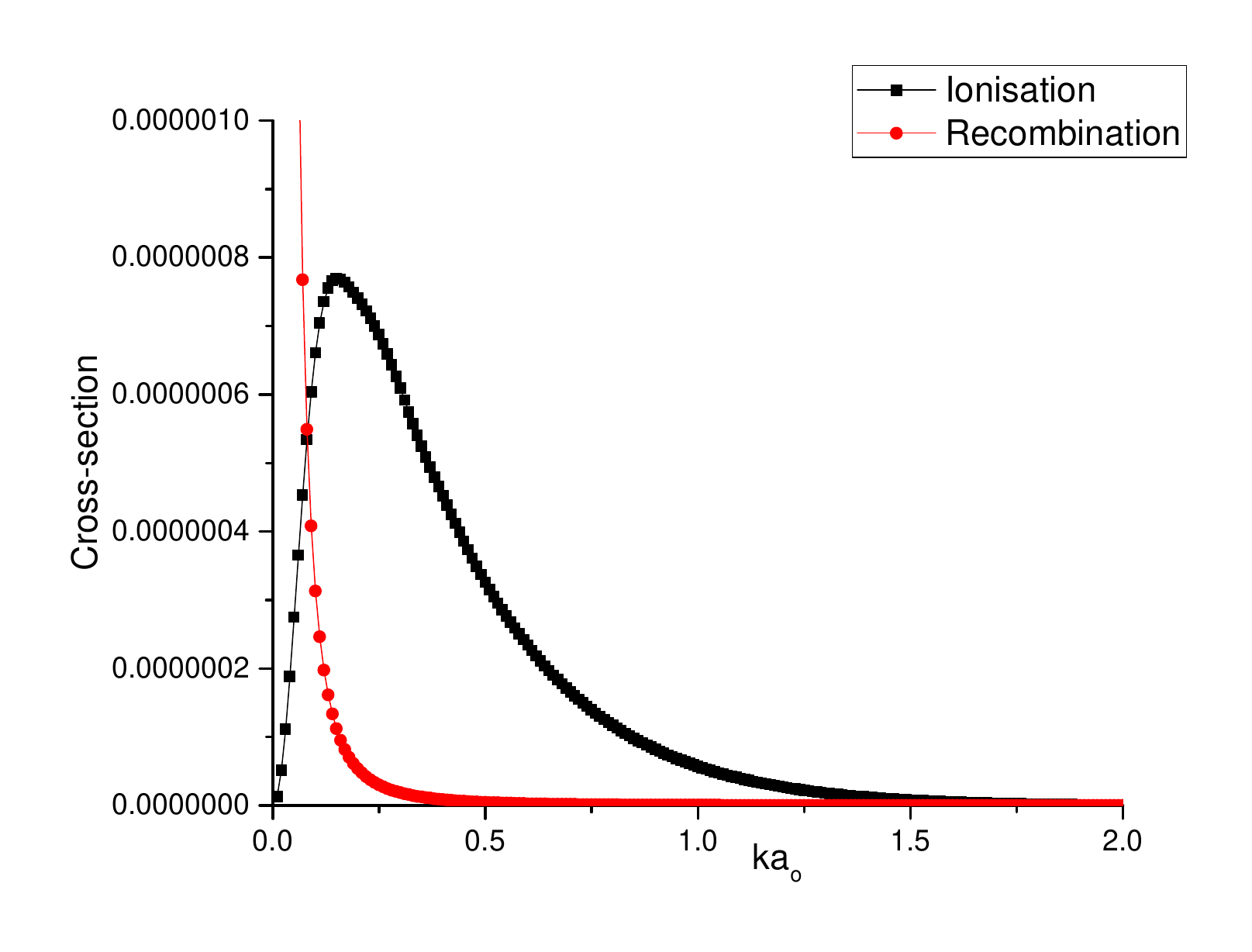}}
	\caption{Total electron impact cross-section for the alkali atoms( Li, Na, K, Rb, Cs) atoms at lower energy range$(\sim 1eV)$ plotted against electron momentum $ka_o$. The cross-section values are calculated for different energy levels of n=20 (squares), 40 (circles) and  60(triangles). For comparison, the cross-section values have been plotted together(bottom-right) for n=60. } \label{cross}
\end{figure} 

In a UCP, the Rydberg atoms formed due to TBR have longer lifetimes, but they suffer electron-atom scattering, due to which they get ionised again after the recombination. This electron impact ionisation is theorised to have caused the spontaneous transition of a beam of Rydberg atoms into a UCP by several experimental groups \cite{kilian2,vanhaecke,robinson}. The existing theoretical explanation for the dissociation employs classical Monte Carlo simulations \cite{robicheaux} and molecular dynamics simulations \cite{Pohl}, which are unable to account for the observations like ultrafast expansion of the UCPs as well as the presence of a minimum energy state for the Rydberg ionisation. Our previous work \cite{prakash} treated this dissociation as a quantum mechanical scattering phenomenon, which gives insights into the underlying atomic processes. For a UCP system, when a Rydberg atom is formed due to recombination, the electron's electric field polarises the highly extended electron cloud of the Rydberg atom, resulting in the polarisation potential. Since the polarisability of alkali Rydberg atoms varies as $n^6$, it becomes a significant factor in the expanding UCP. The polarisation potential for electron-atom interaction in a partially ionised semiclassical plasma has been extensively studied \cite{adams, ramazanov, madina, ramazanov2}. For alkali plasmas, Redmer and Ropke modified the form of the semiclassical potential \cite {redmer} and suggested a screened polarisation potential by adding a screening parameter($\kappa$) to account for the plasma effects. For our UCP system, we have used the screened polarisation potential in the form used by Redmer et. al., which is given by:

%For alkali plasmas, Redmer and Ropke modified the form of the semiclassical potential used by Adams et. al. and  suggested a screened polarisation potential by adding a screening parameter($\kappa$) to account for the plasma effects \cite{redmer}. For our UCP system, we have used the screened polarisation potential in the form used by Redmer et. al., which is given by:

 \begin{equation}
V_{\mathrm{p}}^{\mathrm{s}}(r)=\frac{-e^2\alpha_p}{8 \pi \epsilon_0(r+{r_0})^4}{e^{-2\kappa r}}{(1+\kappa r)}^2
\end{equation} 

Here, $\alpha_p$ is the polarisability, $\kappa$ is the screening parameter and $r_o$ is the cutoff distance parameter. To compute the non-relativistic polarisability $\alpha_p$ for alkali atoms, we use, following Shevelko et. al. \cite{shevelko}, 

$$\alpha_p=\beta_{nl}^{z}= Z^{-4}[n^6+7/4n^4(l^2+l+2)][a_0^3]$$

 Here, Z represents the atomic number, n is the principal quantum number, and l is the angular momentum quantum number. 

\begin{figure}[!h]
 \centerline{ \includegraphics[width=18 cm]{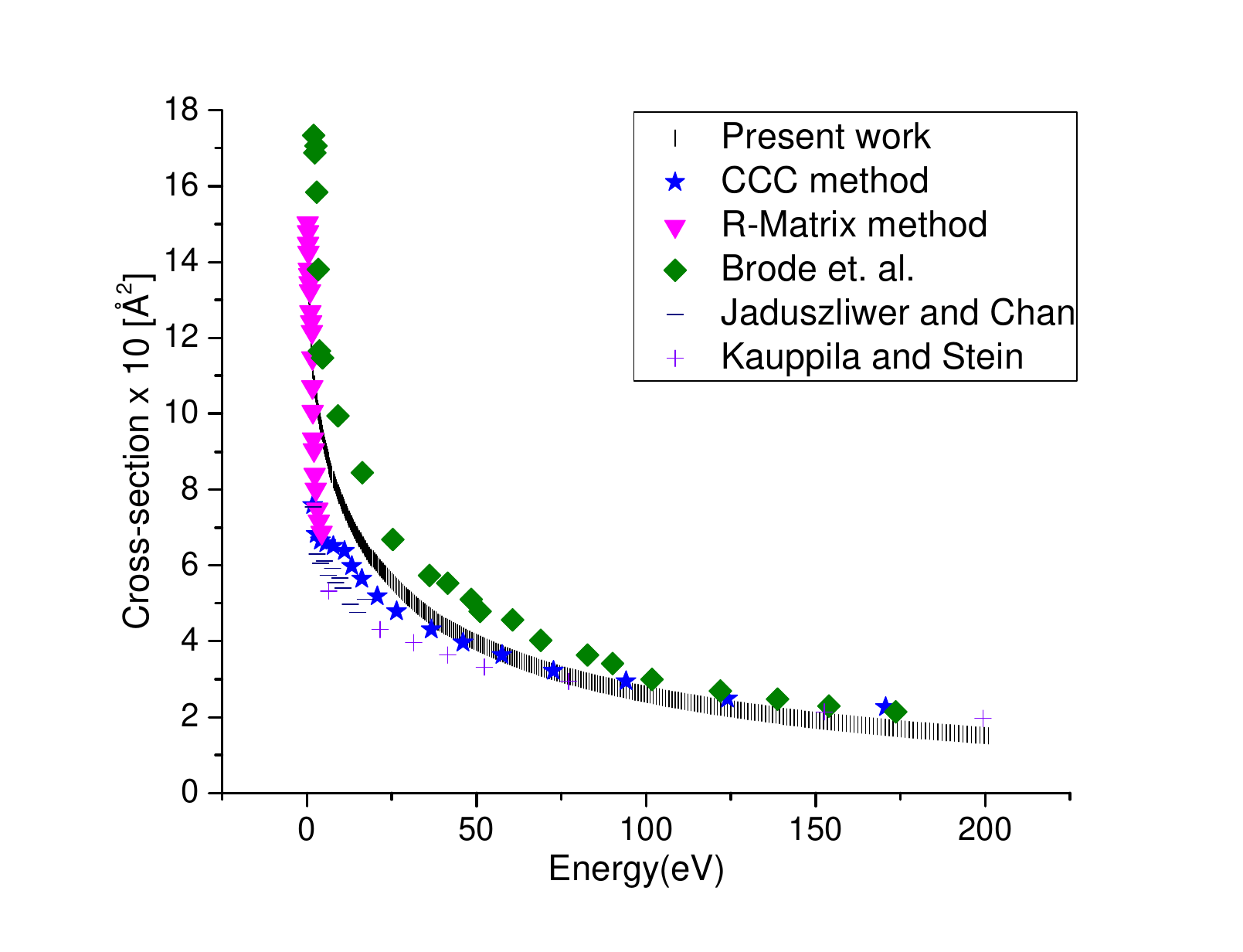}}
  \caption{Unscreened total electron-atom cross-section calculation for Cesium (1–200 eV), shown against Brode (diamonds), Jaduszliwer $\&$ Chan (horizontal bar), Kauppila $\&$ Stein (cross), CCC method(star), and R-matrix method (lower triangle). Our results using the unscreened polarisation potential is shown by Vertical bars\cite{prakash,macaskill, bray, bray2}.}
\label{total_cross_section}
\end{figure}

From above, the potential for electron - atom scattering is computed using partial wave decomposition of the incoming wave as 

\begin{equation}
	\psi_{in}= e^{ikz}=\sum_{l=0}^\infty(2l+1)i^lj_l(kr)P_l(\cos\theta)
\label{partial_in}	
\end{equation}

 The scattered wave function is calculated as the sum of the scattering amplitudes, depending on the phase shift $(\delta_l)$ which arise due to scattering. The phase shift is therefore calculated as the asymptotic value of the phase function $\delta_l(r)$ described by the phase equation \cite{calogero}. 

\begin{equation}
	\delta_l'(r)=-k^{-1}V(r)[\cos\delta_l(r)\hat{j}_l(kr)-\sin\delta_l(kr)\hat{n}_l(kr)]^2; \lim_{r \to \infty} \delta_l(r)=\delta_l
\label{phase_shift}	
\end{equation} 
 
 The calculation of the phase shift as an asymptotic limit is shown in equation(6). Equation(7) gives the calculation of the contribution of each partial wave. The total cross-section is then obtained as the sum of individual partial-wave contributions.

\begin{equation}
	\sigma_l=\frac{4\pi}{k^2}(2l+1)\sin^2\delta_l \hspace{1.5 cm} \sigma_{total}=\sum_{l=0}^\infty \sigma_l
\end{equation}

Figure \ref{cross} shows the electron - atom scattering cross section calculated from above method, for different alkali atoms -  Lithium, Sodium, Potassium, Rubidium and  Cesium in the order, as a function of electron momentum, for Rydberg levels n = 20, 40, and 60 respectively. 
 We have approximated the s-wave cross-section to be the total cross-section as it becomes dominant when the electron energy decreases. This is due to the fact that the centrifugal barrier $V'(r)=(\hbar^2/2\mu)(l(l+1)/r^2)$   becomes significant for higher angular momentum waves, thus impeding them from reaching near the scattering atom. The cross-section values are maximum for low electron momenta and decrease rapidly for higher electron momenta,  due to the higher phase shift that electrons experience at lower momenta. No bound states could be observed, as the highest phase shift we could obtain was 0.2 radian, which is much less than 3.14 radian required for bound state formation as per Levinson's theorem \cite{ma}. An important observation from the graph is that the electron-atom scattering cross-section increases as we go from Cesium to Lithium atom. Hence, it suggests that as the atomic size decreases, the UCP system becomes more sensitive to temperature and density variations.

We have verified our methodology by comparing our potential scattering cross-section calculations with those calculated by Brode et al., Jaduszliwer and Chan, Kaupilla and Stein \cite{macaskill, bray, bray2} for low-temperature plasmas. The cross-section has been calculated for the Cesium atom in the ground state, as shown in figure \ref{total_cross_section}.

This data shows that the experimental and theoretical results are consistent at higher scattering energies (100–200 eV), but noticeable discrepancies appear in the cross-section calculations at lower energies. Our calculated Cesium cross-sections are in agreement with those obtained by Bray and Stelbovics \cite{bray, bray2}, who had employed the Convergent Close-Coupling (CCC) method, demonstrating good consistency with experimental data.

\section{Results and Discussion}

\begin{figure}
\subfigure(a){\includegraphics[width=7.2 cm]{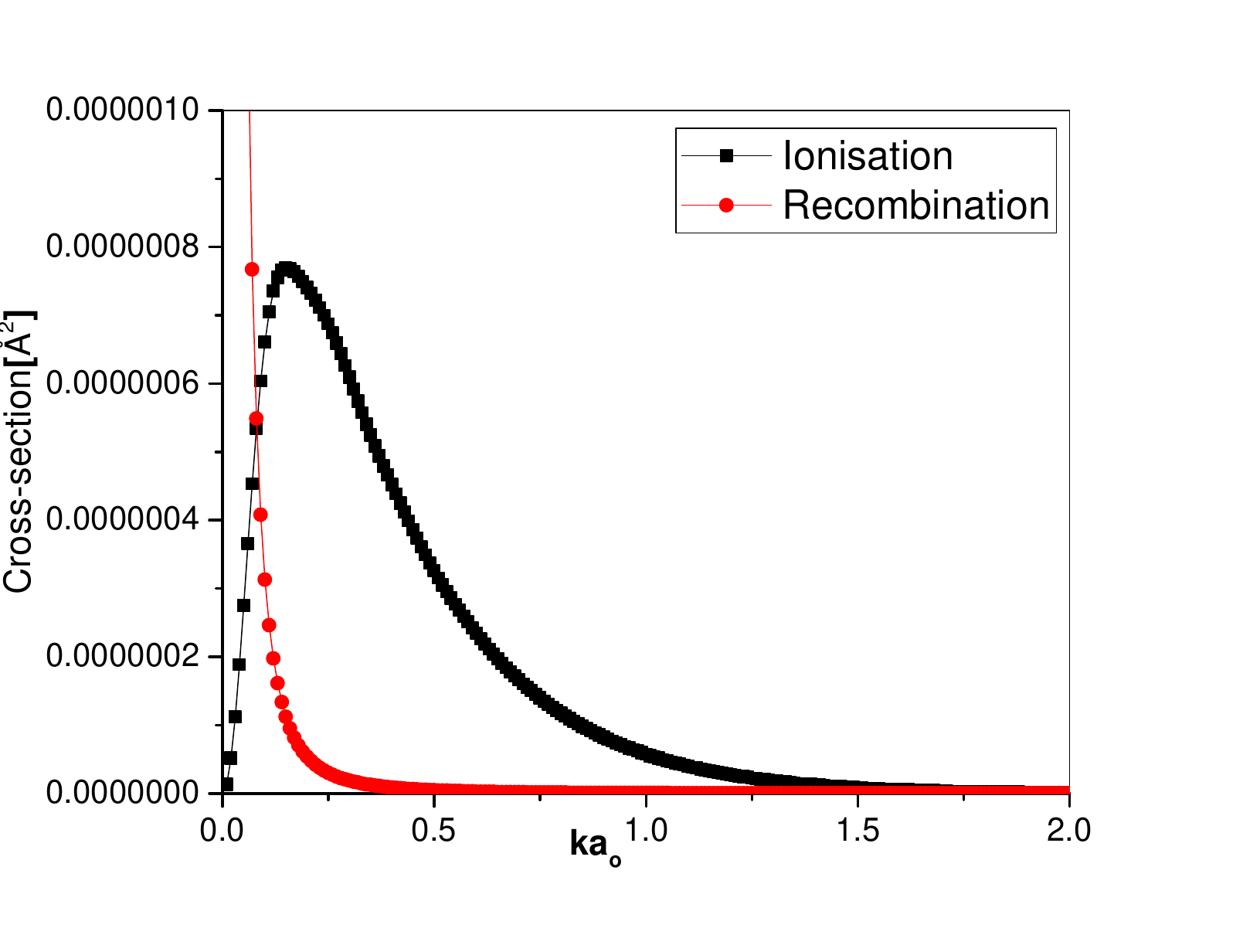}}
\subfigure(b){\includegraphics[width=7.2 cm]{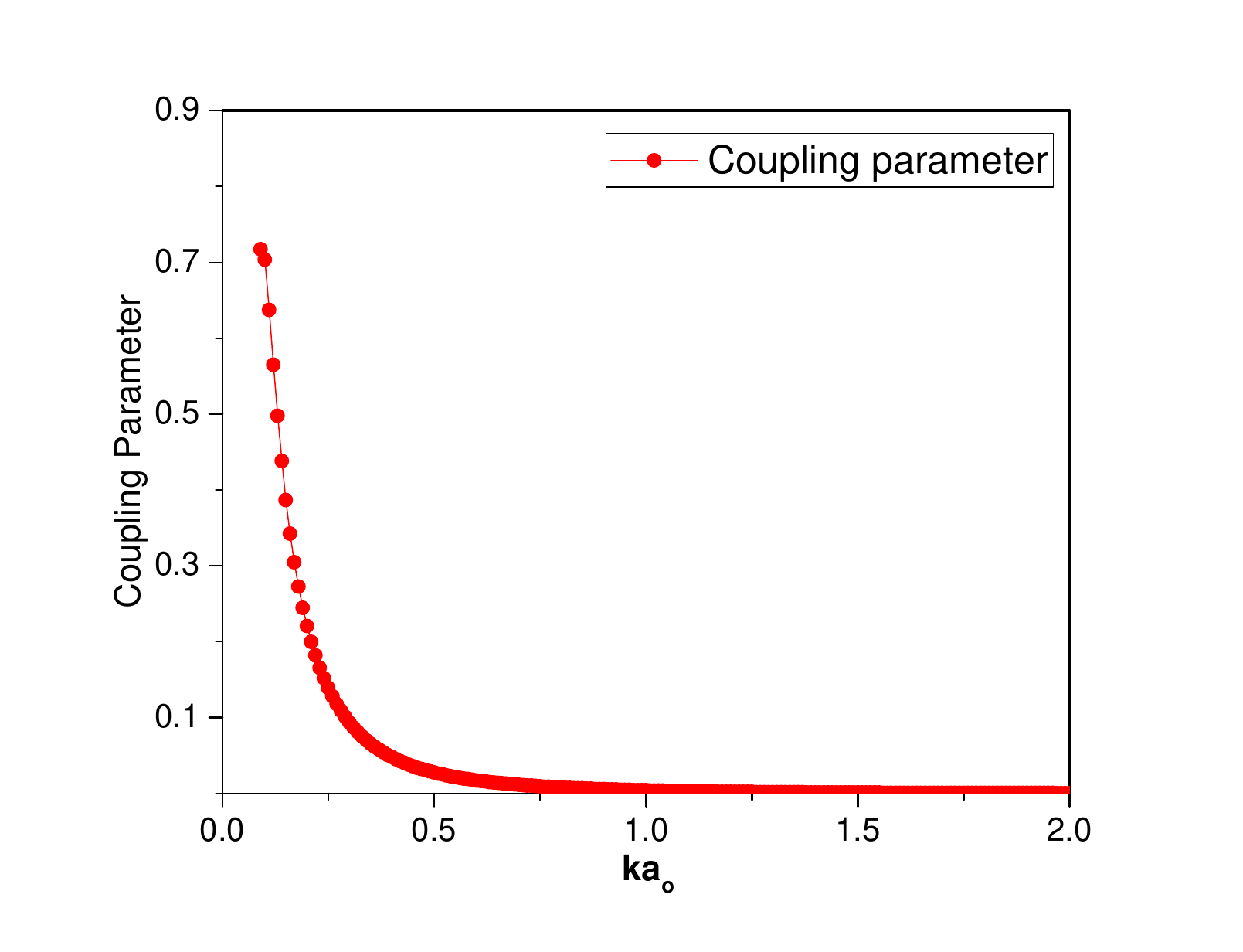}}
\subfigure(c){\includegraphics[width=7.2 cm]{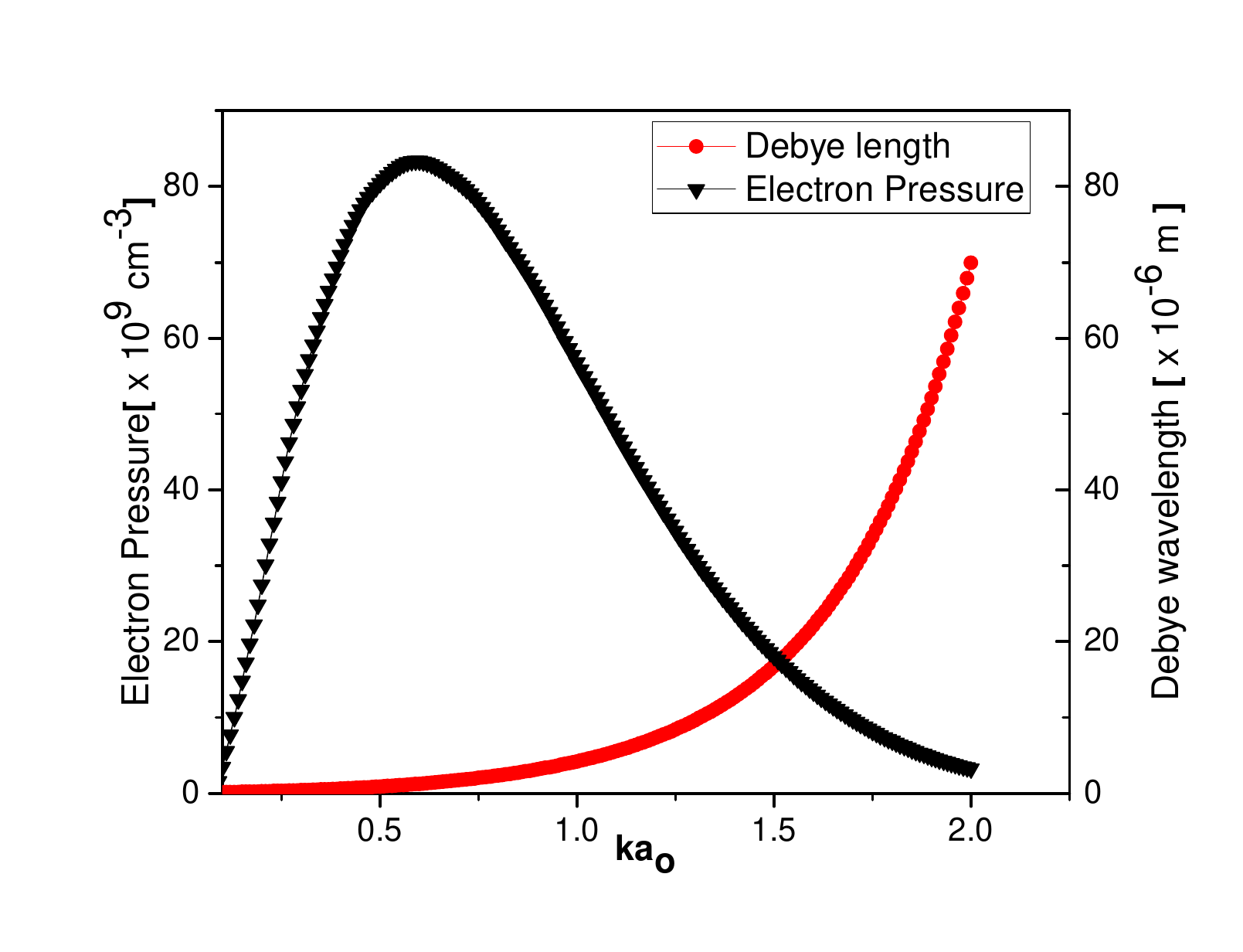}}
\subfigure(d){\includegraphics[width=7.2 cm]{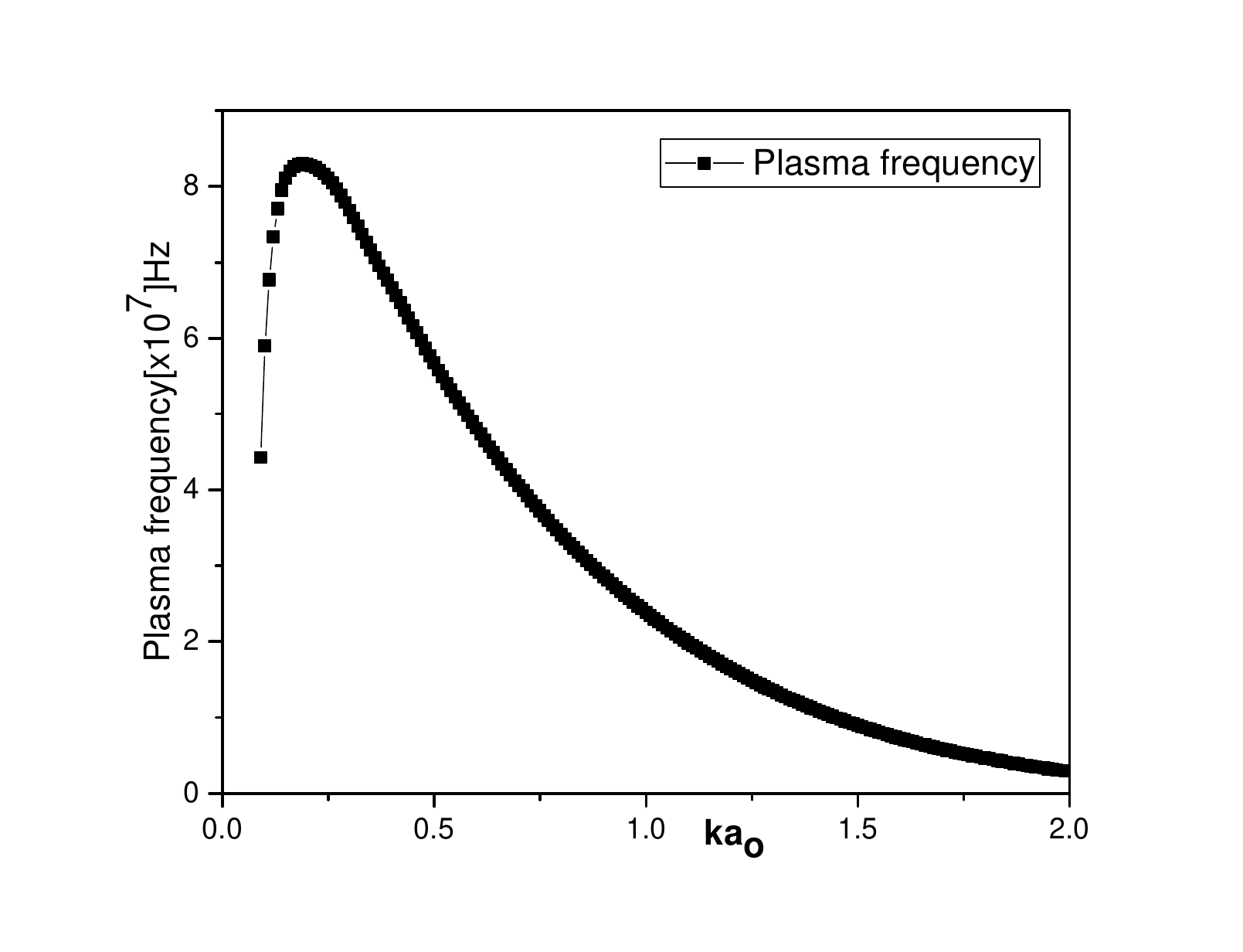}}
\caption{Plasma parameters calculated using the ionisation/recombination rate present in the UCP medium as a function of electron momentum $ka_o$. The graph (a) shows the competing processes of ionisation/recombination in phase space.  Graph (b) shows the coupling parameter.  The graphs (c) shows Debye wavelength and electron pressure, and (d) shows plasma frequency in the phase space.}
\label{plasma_diagnostics}
\end{figure}

When the UCP is in equilibrium, the electron velocities immediately after ionization will follow a Maxwell-Boltzmann distribution $N(v)=(v^2/{v}^3)e^{-v^2/v^2}$. The rate of recombination of these electrons is given by the dominant process of TBR, which is proportional to $T^{-(9/2)}$. Hence, the total rate of recombination is given by

 $$\frac{dN_1}{dt}=\int{v^{-9/2}}N(v)dv=\int{v^{-5/2}}e^{-v^2/v^2}dv$$.

 where the mean number of Rydberg atoms are calculated by integrating the recombination rate. We can also calculate the population of Rydberg levels as the electron populates the Rydberg levels with an energy $E_{n'}=k_BT$.

Now, these Rydberg atoms are reionised due to electron-atom interaction. The total rate of ionisation has been calculated in our previous works, and it is given by $\frac{dN_2}{dt}=\int{N(v)\sigma(v)dv}$. The two rates have been calculated so that the net rate of loss of Rydberg atoms is given by $\frac{dN}{dt}=\frac{dN_2}{dt}-\frac{dN_1}{dt}$ Hence, we have calculated the probability of recombination and the probability of ionisation to calculate the net probability of obtain the Rydberg atom in an alkali UCP in the timescale $\sim$50 $\mu s$.

We calculated parameters of the UCP  from the rates of ionization and recombination. The results are shown in figure \ref{plasma_diagnostics}. It shows that the Debye length increases in space.  This is valid because initially UCP start with very low kinetic energy and then expands rapidly. The electron pressure first increases until the electrons thermalise and then decreases as the temperature and density drop. While the plasma expands, the coupling parameter for electrons decreases due to a density drop. In our case the moderate density of $2 \times 10^5$ cm$^{-3}$ has been taken. The initial density is not sufficient for the appearance of a strongly correlated system, and the plasma is weakly coupled. The plasma frequency is expected to drop as the density decreases, but our graph shows some deviation as it first increases and then decreases. This may be due to the sudden increase in the electron density resulting from the high recombination rate at low temperature. As the electrons expand, the plasma frequency decreases rapidly as expected.

\begin{figure}
	\subfigure(a){\includegraphics[width=7.6cm]{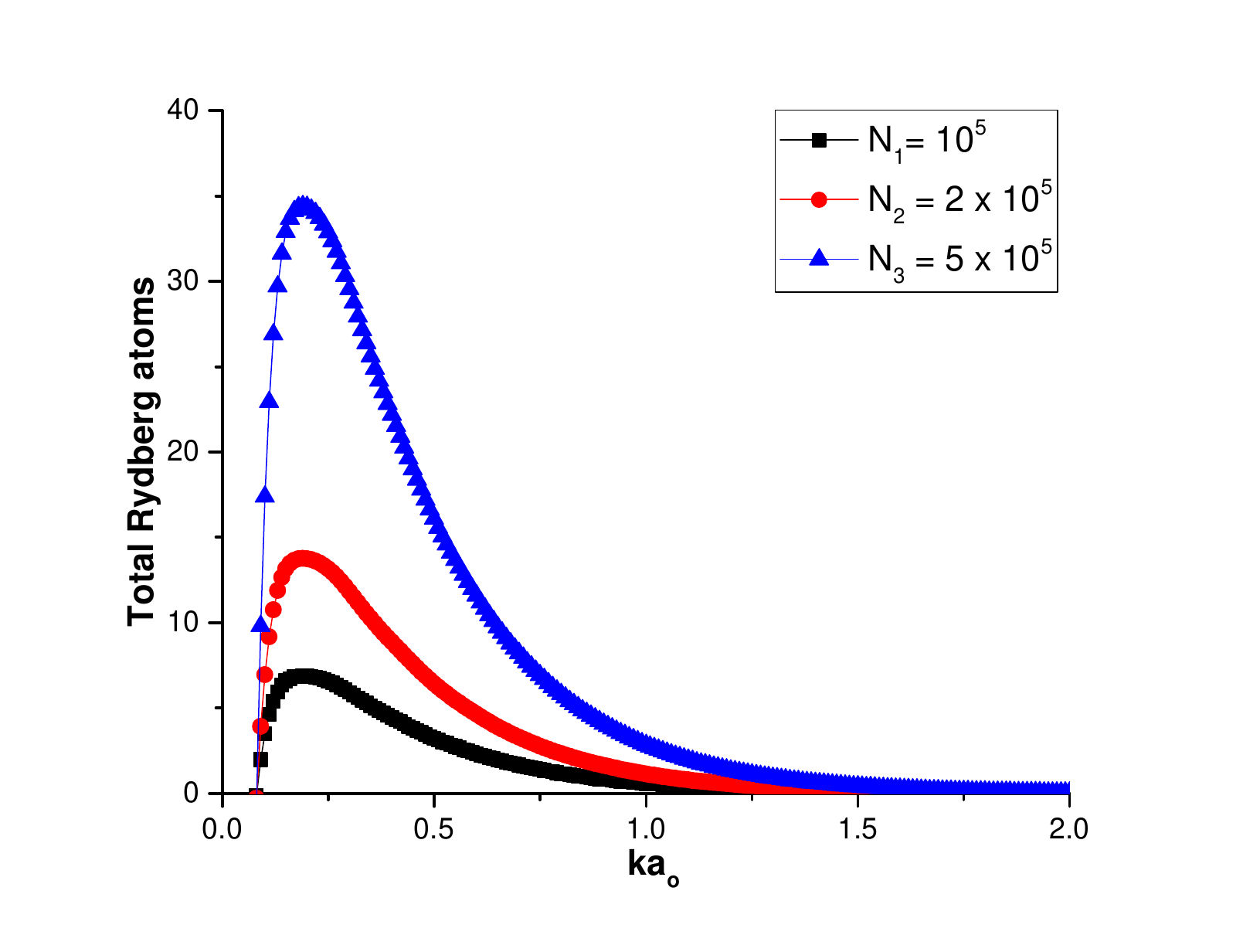}}
	\subfigure(b){\includegraphics[width=7.6cm]{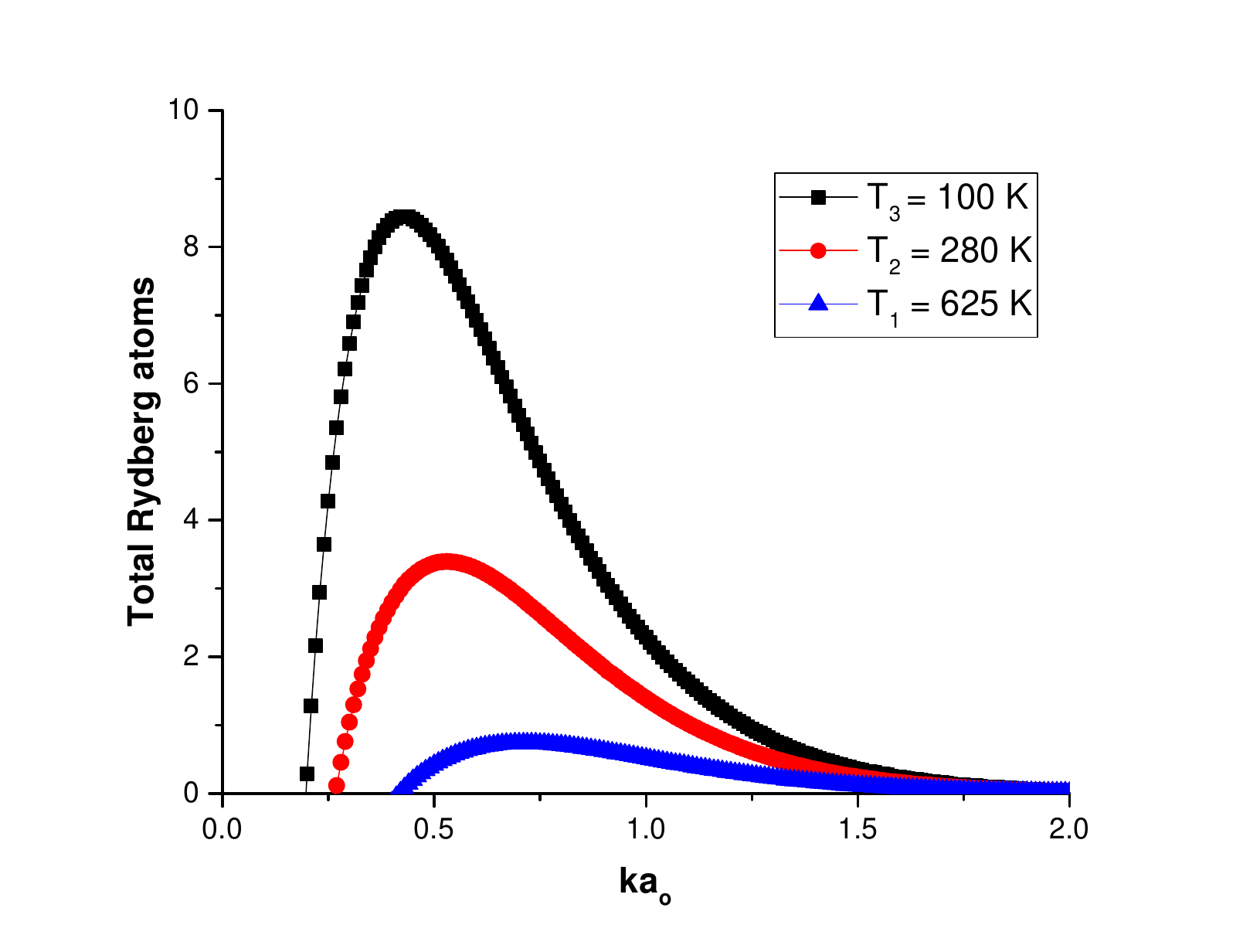}}
	\caption{(a)The upper graph shows total number of Rydberg atoms in the plasma medium as a function of electron momentum. The number has been plotted for different electron densities viz. $10^5, 2 \times 10^5, 5 \times 10^5$( in $cm^{-3})$ for a constant plasma temperature of 100K.g
		(b)The lower graph shows the total number of Rydberg atoms at different temperatures viz. 100K, 278 k, and 625 K. Density has been kept constant at $5 \times 10^5 cm^{-3}$}
	\label{number_of_rydberg_atoms}
\end{figure}

Figure 4(a) is the graph of total Rydberg atoms vs electron momentum. The initial increase in the number, followed by a slow decrease, is in accordance with the experiments of Killian et. al \cite{kilian2}. They had observed deviations from thermodynamic behaviour, which they attributed to three body recombination (TBR). Our calculations show that there exists  a competition between the ionization process and recombination, both of which depend upon interaction of neutral atoms and electrons, which are characterized by cross sections for respective processes. As  cross sections of both ionization and recombination  depend upon both electron temperature as well as ion-densities, the formation of Rydberg atoms will also depend upon ion densities. Hence, the density-dependent maximum is observed in the distribution of the Rydberg atoms in the various energy levels\cite{kilian2}. 

Fig 4(b) shows that the number of Rydberg atoms decreases as the electron temperature increases. Both the ionisation cross-section and the recombination cross-section decrease with an increase in the electron temperature, leading to a smaller overall number of Rydberg atoms at larger electron temperatures. S.X. Hu used a Quantum TBR calculation to show that at lower temperatures, the recombination populates lower n levels compared to the electronic levels predicted by kinetic theory  \cite{hu}. As per the kinetic theory, the electrons populate those energy levels whose energy is given by $E_n \sim k_BT$. However, the quantum TBR \cite{hu} shows that   the electrons form bound states as many as 20 levels lower than the energy level predicted by kinetic theory when the electron's energy is 0.01 eV. This may be significant in UCP dynamics and as the electron temperature drops, the recombination rate increases and the electrons fall to deeply bound states. If the free electrons collide with the deeply bound states, they undergo a superelastic collision, and the colliding electron gains energy, which explains the disorder-induced heating while the plasma expansion starts \cite{robicheaux} . Additionally, Hu's result shows that the recombination drastically shifts to the states with higher angular-momentum(higher l value), specifically when the scattered electron has lower energy. This explains why the Rydberg atoms are long-lived($\sim 100 \mu s$) \cite{killian, kilian2, vanhaecke, gallagher}. These results are significant and have been experimentally verified.

\section{Conclusion}
We have used a quantum mechanical scattering to compute cross sections for ionization as well as a quantum TBR. From the value of these cross sections, we calculated plasma parameters such as  Debye length, electron pressure, coupling parameter, and plasma frequency calculated in the phase space. They follow the experimentally observed results  qualitatively. The 'quantum pressure' arising due to this modifies the expansion of the UCP, which was observed in early UCP experiments \cite{killian, killian2}. We have shown that the underlying atomic processes in a UCP can be accurately understood if we visualise the competing phenomena of ionisation, recombination, and scattering of electrons in the quantum mechanical regime. This approach is complementary to the other methods used such as Monte-Carlo and MD simulations. 

\section{\textbf {Data availability statement}}

All data that support the findings of this study have been included within the article. Any additional data can be made available upon  request to the authors.

\section{\textbf{Acknowledgments}}

The author SP gratefully recognizes the support provided by CSIR-HRDG under the Ministry of Science $\&$ Technology, Govt. of India, for granting the JRF/SRF fellowship (Award No. 09/0414(13706)/2022-EMR-I), which sustained this research throughout its duration.

%\section{References}

%\begin{thebibliography}{9}

%\end{thebibliography}

\end{document}